\documentclass[twocolumn]{autart} 
\usepackage{amsmath}
\usepackage{graphicx}
\usepackage{amsfonts}
\usepackage{algorithm}
\usepackage{algpseudocode}
\usepackage{siunitx}
\usepackage{longtable,tabularx}
\usepackage{booktabs}
\usepackage{xcolor}
\usepackage{balance}
\usepackage{fancybox}
\usepackage{caption}

\algnewcommand{\Inputs}[1]{%
  \State \textbf{Inputs:} 
   \hspace*{0.3em}\parbox[t]{\linewidth}{\raggedright #1}
}
\algnewcommand{\Initialize}[1]{%
  \State \textbf{Initialize:}
   \hspace*{0.3em}\parbox[t]{.8\linewidth}{\raggedright #1}
}
\algnewcommand{\Output}[1]{%
  \State \textbf{Outputs:}
   \hspace*{0.3em}\parbox[t]{.8\linewidth}{\raggedright #1}
}
\def\p(#1|#2){p\left(#1\,|\,#2\right)}
\def\q(#1|#2){q\left(#1\,|\,#2\right)}

\begin{document}
\begin{frontmatter}
\title{Implicit Particle Filtering {\em via} a Bank of Nonlinear Kalman Filters} \vspace{-22pt}
\author[KU]{Iman Askari}\ead{askari@ku.edu},
\author[Army]{Mulugeta A. Haile}\ead{mulugeta.a.haile.civ@mail.mil},
\author[KU-math]{Xuemin Tu}\ead{xuemin@ku.edu},
\author[KU]{Huazhen Fang}\ead{fang@ku.edu}
\address[KU]{Information \& Smart Systems Laboratory, Department of Mechanical Engineering, University of Kansas
}
\address[Army]{Vehicle Technology Directorate, U.S. Army Research Laboratory
}
\address[KU-math]{Department of Mathematics, University of Kansas
}
 
\begin{keyword}
Nonlinear state estimation;  particle filter; implicit particle filter
\end{keyword}
\begin{abstract}
The implicit particle filter seeks to mitigate particle degeneracy by identifying particles in the target distribution's high-probability regions. This study is motivated by the need to enhance computational tractability in implementing this approach. We investigate the connection of the particle update step in the implicit particle filter with that of the Kalman filter and then formulate a novel realization of the implicit particle filter based on a bank of nonlinear Kalman filters. This realization is more amenable and efficient computationally.
\end{abstract}
\end{frontmatter}

\section{Introduction} 
\indent State estimation has found significant application in all areas of science and engineering to accelerate technological advancements. The particle filtering (PF) approach has arisen as a crucial means, especially for nonlinear non-Gaussian systems. At the core, PF updates an ensemble of particles {\em via} sequential importance sampling such that their empirical distribution approximates the target distribution of the system's state. However, the PF often suffers from particle degeneracy, whereby a majority of the particles see their weights diminish to almost zero, leading to a poor approximation of the target distribution. Even though one can use a large number of particles so that at least some of them have significant weights, the number of particles needed often grows catastrophically with the system's dimension.

A survey of the literature indicates an ongoing search for principled ways to address the above issue. A straightforward, popular method is resampling, which randomly replaces low-weight particles with high-weight ones~\cite{Sarkka:Cambridge:2013}. More sophisticated mechanisms have also been developed, e.g.,~\cite{Chorin:PNAS:2009,Yang:TAC:2013,Raihan:AUTO:2018,Stano:AUTO:2013}. Among them, the method of implicit sampling has inspired much attention. Implicit sampling exploits the idea that fewer particles will be needed as long as they lie in the high-probability regions of the target distribution~\cite{Chorin:PNAS:2009}. By design, it constructs probability distributions assumed for the particles and utilizes them as a reference to select particles from the target distribution's high-probability regions. The resultant implicit PF (IPF) shows effectiveness in keeping the number of particles manageable even for high-dimensional systems~\cite{Chorin:CAMCS:2010,Chorin:Springer:2013}. However, implementing IPF is non-trivial as it entails solving a nonlinear optimization problem and then a nonlinear algebraic equation. While different numerical approaches are used  in~\cite{Chorin:CAMCS:2010,Chorin:Springer:2013,Su:SIAM:2017,Weir:NPG:2013}, the computation is expensive and tedious. This motivates the development of efficient alternative approaches to approximate the IPF.

This paper proposes that the IPF framework can be approximately realized by a bank of nonlinear Kalman filters (KFs). This realization shows more computational tractability while maintaining the IPF's merit to sample from regions of high probability to achieve high estimation performance. Centering around this insight, we first introduce implicit importance sampling. This will provide a basis to understand the IPF and a link to connect the IPF with the KFs. Then, we synthesize the approach of using a bank of nonlinear KFs to approximately execute the IPF. Here, we particularly use the extended and unscented KF (EKF and UKF, respectively), as they are among the most commonly used KFs. Finally, we discuss the connections between the proposed IPF realization with several existing PF methods and validate our approach through simulation.


\section{Implicit Importance Sampling}\label{IIS}
In general, a Bayesian estimation problem involves computation of the conditional expectation of an arbitrary function as follows:
\begin{align}\label{Integral-Bayesian-estimation}
\mathbb{E} \left[g(x) \, | \, y_{1:T}\right] = \int g(x) \p(x | y_{1:T})dx,
\end{align}
where $x \in \mathbb{R}^n$, $g: \mathbb{R}^n \rightarrow \mathbb{R}^m$ is an arbitrary function, and $\p(x | y_{1:T})$ is the posterior probability density function (PDF) of $x$ conditioned on discrete measurements $y_{1:T} = \left\{ y_1, y_2, \ldots, y_T \right\}$~\cite{Sarkka:Cambridge:2013}. The above integral often prohibits  closed-form evaluation for a nonlinear $g$. As a remedy, the Monte Carlo method approximates $\p(x | y_{1:T})$ by a set of particles (samples) and then computes the integral in~\eqref{Integral-Bayesian-estimation} based on the particles. Yet, drawing particles directly from $\p(x | y_{1:T})$ can be challenging.  
Importance sampling thus proposes to sample from a different distribution $q(x)$, called importance distribution, and weight the particles accordingly. However, it may require an enormous number of particles to achieve just fair accuracy, because the particles  taken from $q(x)$ may fall into low-probability regions of $\p( x | y_{1:T})$. To ameliorate this problem, a promising approach  is {\em implicit importance sampling}, as introduced below.

 Consider the  task of drawing particles $ x^i$ for $ i=1,\ldots,N $ from high-probability regions of  $\p(x | y_{1:T})$. To this end,  we can  introduce a reference random  vector $\xi$ with a known PDF $p(\xi)$ to  create a map $\xi \rightarrow x^i$ that aligns the target $\p(x^i | y_{1:T})$ with reference $p(\xi)$ density. This notion delivers a two-fold benefit. First, the map will allow to connect high-probability samples of $\xi$ to highly probable $x^i$. Second, one can choose $p(\xi)$ such that it is easy to sample, thus reducing the difficulty of sampling  $x^i$. 
Proceeding forward, write $X$ as a shorthand  for $x^i$, and define $F  (X) = - \log \left(\p(X | y_{1:T})  \right)  $ and $s(\xi) =-\log \left ( p(\xi) \right) $. The highest probabilities of $\p(X | y_{1:T})$ and $p(\xi)$ will appear at points where $\min F(X)$ and $\min s(\xi)$ are achieved, respectively. Then, to align $\p(X | y_{1:T})$ with $p(\xi)$ around each other's highest probability point, we can let
\begin{align}\label{map-xi-X}
F  (X) - \min F   (X) = s (\xi) - \min s (\xi),
\end{align}
which implies a map $X(\xi)$. With $X(\xi)$, one can draw a particle from the high-probability regions of $\xi$ and map it to obtain a particle of $X$, which is ensured to lie in the high-probability regions of $\p(X | y_{1:T})$. Here, a trick to pick a high-probable particle for $\xi$ is by using an auxiliary probability distribution that has the same support but concentrates on the high-probability regions of $p(\xi)$. 

Next is deciding the sampling weight for $X$. Based on the above, the importance distribution $\pi(\cdot)$ is the PDF of $X(\xi)$ as a mapping of $\xi$ given $p(\xi)$. Assume that $X(\xi)$ is invertible for simplicity of exposition (the result is readily generalizable to the non-invertible case based on~\cite[Corollary 11.3]{Jacod:Springer:2004}). Then, by~\cite[Corollary 11.2]{Jacod:Springer:2004}, it follows that
\begin{align*}
 \pi( X(\xi) )  = \left| J \right|^{-1} \cdot p(\xi),
\end{align*}
where $\pi(X)$ denotes the importance distribution of $X$,   $J = dX(\xi) / d \xi$, and $\left| J\right|$ is the absolute value of J's determinant. The weight of $X$ using~\eqref{map-xi-X} hence is
\begin{align} \label{w_tilde_IIP_compute} \nonumber
\tilde w &= \frac{ \p( X | y_{1:T} ) }{ \pi (X(\xi)) } = \frac{\exp(-F  (X))}{|J|^{-1}\cdot p(\xi)}\\
&= |J | \cdot \exp \left[-\min F  (X)+\min s(\xi ) \right].
\end{align}
For $x^i$, the  normalized weight  is 
then given by
\begin{align}\label{w_IIP_compute}
w^i = \frac{\tilde{w}^i}{\sum_{j=1}^{N} \tilde{w}^j} = \frac{|J^i| }{\sum_{j=1}^N |J^j| }.
\end{align}
As a result,  \eqref{Integral-Bayesian-estimation}  is approximated as
\vspace{-5pt}
\begin{align}\label{integral-eval}
\mathbb{E}\left[g(x) \, | \, y_{1:T}\right] \approx \sum_{i=1}^N w^ig(x^i).
\end{align}

In above,~\eqref{map-xi-X}-\eqref{integral-eval} comprise  the method of implicit importance sampling. More details are offered in~\cite{Atkins:MWR:2013,Morzfield:Comms:2015}. While this method promises to use fewer particles, it is computationally challenging to find $\min F(X)$ and then solve \eqref{map-xi-X}, due to the often nonlinear non-convex $F$. To find an alternative way forward, we examine a Gaussian special case here  to gain an insight into tackling this issue.

\noindent\rule{\columnwidth}{1pt}

\noindent{\bf Gaussian Case Study}: Consider the following Gaussian approximation:
\begin{align*}
 p\left(\left[
\begin{matrix}
x \cr y_{1:T}
\end{matrix}
\right] \right) \sim 
\mathcal{N}\left(  
\left[
\begin{matrix}
\bar m \cr \bar y_{1:T}
\end{matrix}
\right],
\left[
\begin{matrix}
P^x & P^{xy} \cr (P^{xy})^\top & P^y
\end{matrix}
\right]
\right).
\end{align*}
It then follows that
$
 \p( x   |   y_{1:T}) =  \mathcal{N} ( \tilde m, \tilde P^x)$,
where
\vspace{-5pt}
\begin{align*}
\tilde m &=  \bar m  + P^{xy} \left( P^y \right)^{-1} \left( y_{1:T} - \bar y_{1:T} \right),\\
\tilde P^x &= P^x - P^{xy} \left( P^y \right)^{-1} \left( P^{xy}\right)^\top.
\end{align*}
Suppose   $p(\xi) \sim  \mathcal{N} ( 0, I)$ without loss of generality. The map $X(\xi)$ derived from~\eqref{map-xi-X}  is  
\vspace{-5pt}
\begin{align*}
X(\xi) = \tilde m +\sqrt{ \tilde P^x }  \xi.
\end{align*}
After a high-probability particle $\xi^i$ is drawn from $p(\xi)$,   it can be mapped by $X(\xi)$ to compute   the particle $x^i$. Its weight  can be determined using~\eqref{w_tilde_IIP_compute}-\eqref{w_IIP_compute}. Specifically,
\begin{align*}
x^i = \tilde m + \sqrt{ \tilde P^x } \xi^i, \ \ 
w^i = {1 \over N}.
\end{align*}
Based on~\eqref{integral-eval}, we have
\vspace{-10pt}
\begin{align*}
{\mathbb E} \left[ g(x) \, | \, y_{1:T} \right] \approx  {1 \over N} \sum_{i=1}^N g(x^i).
\end{align*}

\vspace{-20pt}

\noindent\rule{\columnwidth}{1pt}

The  above case study indicates that implicit importance sampling is efficiently executable in a Gaussian setting. Extending this insight, we identify that, under the Gaussian approximation, the IPF can be approximately implemented as a bank of KFs.

\section{Approximation of IPF as a Bank of KFs}

Consider the problem of estimating the state of a nonlinear system taking the form
\begin{align}\label{nonlinear-state-space-sys}
\left\{
\begin{aligned}
      x_{k+1} &= f(x_{k}) +w_{k},\\
      y_{k} &= h(x_{k}) +v_{k},\\
\end{aligned}      
\right.
\end{align}
where $x_{k} \in \mathbb{R}^{n_x}$ is the state,  $y_{k} \in \mathbb{R}^{n_y}$ is the measurement, and  $w_{k} $ and $v_{k}$ are  zero-mean white noises with covariances $Q$ and $R$, respectively.  For the purpose of state estimation, it is of interest to consider the conditional PDF $\p(x_{0:k} | y_{1:k} )$.  By the Markovian property of  \eqref{nonlinear-state-space-sys},   $\p(x_{0:k} | y_{1:k} )$ satisfies the following recursive relation:
\begin{align*}
\p(x_{0:k}|y_{1:k}) \propto \p(y_{k}|x_{k}) \p(x_{k}|x_{k-1}) \p(x_{0:k-1}|y_{1:k-1}). 
\end{align*} 
One can assume that   $\p(x_{0:k-1} | y_{1:k-1})$ has been made  available at time $k-1$, as a result of the preceding recursive updates, and that it can be weakly approximated by the empirical distribution of  particles  $ x_{0:k-1}^i$ for $i=1, \ldots, N  $ with   weights $w_{k-1}^i$.  Therefore, to draw   new particles $x_k^i$ at time $k$, we    only need to consider 
\begin{align}\label{IPF_recursive_update}
\p(x_{k}^{i}|y_{1:k}) \propto \p(y_{k}|x_{k}^{i}) \p(x_{k}^{i}|x_{k-1}^{i}). 
\end{align}
Using $X_k$ as a shorthand for $x_k^i$, we define  
\begin{align*}
F_i(X_k)  
&= - \log\left( \p(y_{k}| X_k ) \p(X_k |x_{k-1}^{i}) \right),
\end{align*}
To pick $X_k$ in high-probability regions, we  choose  a reference random vector $\xi_k$ with $p(\xi_k)=\exp(-s(\xi_k))$, and following  \eqref{map-xi-X}, let
\begin{align}\label{map-xi-X-PF}
F_i(X_k) - \min F_i(X_k) = s (\xi_k)  -\min s (\xi_k ). 
\end{align}
Taking a high-probability sample $\xi_k^i$ from $p(\xi_k)$ and   solving~\eqref{map-xi-X-PF}, we can obtain  a desired particle $x_k^i$, with weight  
\begin{align} \label{w_tilde_compute} \nonumber
\tilde {w}_{k}^i  &=   {w}_{k-1}^i   \frac{\p(y_k|X_k ) \p(X_k |x_{k-1}^i)}{\pi( X_k (\xi_k) )}  \\
&=  {w}_{k-1}^i    |J_k^i | \cdot  \exp \left[- \min{F_i(X_k)} + \min{s (\xi_k)} \right] ,
\end{align} 
where $J_k^i = d X_k (\xi_k^i) / d \xi_k^i$. 
The normalized weight then is
\begin{align}\label{weight-update}
w_k^i =    \frac{\tilde{w}_k^i}{\sum_{j=1}^{N} \tilde{w}_k^j}.  
\end{align}
Here,~\eqref{map-xi-X-PF}-\eqref{weight-update} illustrate the particle computation and weighting underlying the IPF. However, a difficulty for the implementation lies in finding $\min F_i(X_k)$ and solving the algebraic equation~\eqref{map-xi-X-PF}. Even though the literature presents several useful methods based on iterative computation~\cite{Chorin:Springer:2013}, their computational complexity is still not-trivial. To address the challenge, we   take inspiration from the Gaussian Case Study in Section~\ref{IIS} to  solve~\eqref{map-xi-X-PF}. 

First, let us   introduce a   Gaussian approximation:
\begin{align}\label{Gaussian_assumption_IPF_KF}
p \left(\left. \begin{bmatrix}
x_k^i \\ y_{k} 
\end{bmatrix} \, \right| \, x_{k-1}^i\right) \sim \mathcal{N}\left(\begin{bmatrix}
\bar{m}_k^i \\ \bar{y}_k^i 
\end{bmatrix}, \begin{bmatrix}
\bar{P}_k ^{i} & \bar{P}_k^{xy,i} \\  \left(\bar{P}_k^{xy,i} \right)^\top & \bar{P}_k ^{y,i}
\end{bmatrix}\right).
\end{align}
By~\eqref{Gaussian_assumption_IPF_KF}, we have
\begin{align}\label{Particle_update}
\p( x_k^i  |  y_k, x_{k-1}^i ) \sim \mathcal{N} \left(  \tilde m_k^i , \tilde P_k^i  \right), 
\end{align}
where
\begin{subequations}\label{KF-update}
\begin{align}\label{KF-update-a}
\tilde m_k^i  &= \bar m_k^i +  \bar{P}_k^{xy,i} \left( \bar{P}_k ^{y,i} \right)^{-1} \left( y_k - \bar y_k^i \right)   ,\\ \label{KF-update-b}
\tilde P_k^i &= \bar{P}_k ^{i}  - \bar{P}_k^{xy,i} \left( \bar{P}_k ^{y,i} \right)^{-1} \left(\bar{P}_k^{xy,i} \right)^\top.
\end{align}
\end{subequations}
Assuming $ \p(y_k|x_k^i) = \p(y_k|x_k^i,x_{k-1}^i) $ and inserting it into~\eqref{IPF_recursive_update}, we obtain
\begin{align}\label{IPF_recursive_update_new}
\p(x_{k}^{i}|y_{1:k}) \propto     \p(x_k^i|y_k,x_{k-1}^i) \p(y_k|x_{k-1}^i).
\end{align}
Given~\eqref{IPF_recursive_update_new}, we use~\eqref{map-xi-X-PF}  to align  $\p(x_{k}^{i}|y_{1:k})$ with  the high-probability regions of  a standard Gaussian random vector  $\xi_k$, with $p(\xi_k) \sim \mathcal{N}(0,I)$.  Then,  referring to the Gaussian Case Study in Section~\ref{IIS}, 
the desired particle $x_k^i$   can be expressed as
\begin{align}\label{particle-compute}
x_k^i = \tilde m_k^i +  \sqrt{\tilde{P}_{k}^{i}}\xi_k^i,
\end{align}

\vspace{35pt}

\noindent\rule{\columnwidth}{1pt}

\vspace{-10pt}

\noindent{\bf Algorithm 1: The {KF-IPF} Algorithm}
\vspace{-5pt}
\begin{itemize}

\item Initialize a set of particles $x_0^i$ for $i=1,\ldots,N$ with   $w_0^i$ and $P_0^i$ at $k=0$
\item Implement the following for $x_k^i$ for $i=1,\ldots,N$   from $k=1, \ldots,T$

\begin{itemize}

\item[-] Compute $\bar m_k^i$ and $\bar P_k^i$ using $x_{k-1}^i$ and $P_{k-1}^i$ {\em via} KF prediction

\item[-] Compute $\bar y_k^i$, $\bar P_k^{y,i}$ and $\bar P_k^{xy,i}$ as in KF prediction

\item[-] Compute $\tilde m_k^i$ and $\tilde P_k^i$  {\em via} KF update in~\eqref{KF-update-a}-\eqref{KF-update-b}

\item[-] Compute $x_k^i$ and $w_k^i$  {\em via}~\eqref{particle-compute}-\eqref{Cov_perturb}

\item[-] Do resampling if necessary

\item[-] Output the state estimate

\item[-] $k \leftarrow k+1$

\end{itemize}

\end{itemize}

\vspace{-15pt}

\noindent\rule{\columnwidth}{1pt}

\noindent where $\xi_k^i$ is a high-probability sample of $p(\xi_k)$. 
Based on~\eqref{w_tilde_compute}-\eqref{weight-update}, its weight is computed as
\vspace{-2pt}
 \begin{align}  \label{weight-compute}
{w}_k^i &= \frac {w_{k-1}^i \p(y_k|x_{k-1}^i)  }{\sum_{j=1}^N  w_{k-1}^j  \p(y_k|x_{k-1}^j)}. 
 \end{align}  

In practice, it is important that $\xi_k^i$ are drawn from the high probability regions of $p(\xi_k)$. An implementation trick to ensure this is to sample from $\mathcal{N}(0, \alpha I )$ with $0< \alpha \ll 1$. We emphasize that $\alpha$ can be considered  as a tunable parameter to dictate  the desired high probability region of $p(\xi_k)\sim\mathcal{N}(0, I)$. It hence must take a small enough value to ensure that only highly probable particles with respect to $\mathcal{N}(0, I)$ are drawn. This is especially true when only a small number of particles are used. The covariance associated to $x_k^i$ is given as
\vspace{-5pt}

\begin{equation}\label{Cov_perturb}
P_k^i = \tilde{P}_k^i .
\end{equation}

Note that the  formulation in~\eqref{Gaussian_assumption_IPF_KF}-\eqref{KF-update}   is equivalent to the well-known KF update procedure. This connection suggests the viability of  approximately implementing the IPF as a bank of parallel KFs applied to individual particles. 
Specifically, given $x_{k-1}^i$ at time $k-1$,  one can compute $\bar m_k^i$ and $\bar P_k^i$ {\em via} KF prediction and then compute $\tilde m_k^i$ and $\tilde P_k^i$ {\em via} KF update~\cite{Fang:JAS:2018}; going further, $\tilde m_k^i$ combines with $\xi_k^i$ to generate $x_k^i$. Different nonlinear KFs can be leveraged to enable the   prediction-update   of the particles. Here, we highlight the use of the EKF and UKF to approximate the mean and covariance statistics of the Gaussian approximation in~\eqref{Gaussian_assumption_IPF_KF}. The EKF and UKF have found wide use with proven effectiveness for various  systems, and their detailed equations can be found in~\cite{Fang:JAS:2018}. As extensions of the standard linear KF, both EKF and UKF at their core aim to track the conditional statistics (mean and covariance) of a nonlinear system's state recursively using accumulating measurement data. To achieve this end,   EKF linearizes the system's nonlinear functions, and UKF leverages the unscented transform.  

Summarizing the above, Algorithm 1 outlines the KF-IPF algorithm  framework. Based on this framework, 
approximate realizations based on the EKF or UKF can be readily derived and named as E-IPF and U-IPF, respectively. The details are omitted here for the sake of space.

The proposed KF-IPF algorithm provides an amenable way to implement the IPF by approximating it as a bank of KFs. The algorithm circumvents the need to numerically find $\min F_i(X_k)$ and solve the nonlinear algebraic equation~\eqref{map-xi-X-PF} with computationally intensive numerical schemes in existing IPF literature. By design, the KF-IPF can also exploit the advantages of a nonlinear KF. For instance, it can achieve derivative-free computation if the UKF is used. These features will enable it to facilitate the computation of the IPF considerably. Further, the algorithm inherits the capability of the IPF in mitigating the issue of particle degeneracy. Also, note that the development of the KF-IPF algorithm uses the Gaussian approximation in~\eqref{Gaussian_assumption_IPF_KF}, which plays a utilitarian role in solving~\eqref{map-xi-X-PF} {\em via} the KF update. In effect, the approximation serves to facilitate computation. Despite it, the KF-IPF algorithm, as an IPF method, is still well applicable to non-Gaussian estimation problems. The simulation study in Section~\ref{num-sim} further highlights this.

The literature also includes PF methods based on banks of EKF and UKF~\cite{Doucet:SC:2000,Merwe:NIPS:2001}, referred to as EPF or UPF, respectively. However, they are different from the proposed KF-IPF algorithm on several aspects. First, the EPF/UPF is designed to use a bank of EKF/UKF to approximate PF's optimal importance distribution, $\p(x_k | x_{k-1}, y_k)$. However, the KF-IPF is based on the notion of {\em implicit importance sampling}, which looks for high-probability regions of the target distribution and then locates particles in these regions. Second, in terms of implementation, in the sampling step~\eqref{particle-compute}, the EPF/UPF draws particles from the Gaussian approximation made by the EKF/UKF, which is analogous to the case where the KF-IPF directly draws $\xi_k^i$ from $\mathcal{N}(0, I)$. However, a significant risk, in this case, is that $x_k^i$ will fall in the low-probability regions when a marginally probable $\xi_k^i$ is taken---this is also likely why the EPF and UPF sometimes fail to perform adequately, as noticed in~\cite{Sarkka:Cambridge:2013}. By contrast, the actual implementation of the KF-IPF requires focusing particles only on the high-probability regions. For instance, to ensure this, we suggested drawing particles $\xi_k^i$ from $\mathcal{N}(0,\alpha I)$ with $0<\alpha \ll 1$ so that the resultant $x_k^i$ is highly probable. Finally, in the particle weighting step~\eqref{weight-compute}, the KF-IPF utilizes $\p(y_k | x_{k-1}^i)$ instead of $\p(y_k | x_{k}^i)$ as in the case of the EPF/UPF. To sum up, the KF-IPF is distinct from the EPF/UPF, while contributing a new insight into the broad notion of using KF banks to implement the PF.

\section{Numerical Simulation}\label{num-sim}

In this section, we assess the performance of the KF-IPF algorithm on the 40-dimensional Lorenz'96 model~\cite{lorenz96}:
\begin{align}\label{Lorenz96}
 \dot x_j &= f(x, F) = \big(x_{j+1} -x_{j-2} \big)x_{j-1} - x_j +F,
\end{align}
where $j = 1,\hdots,n_x = 40$ is the dimension index,  $x_{-1} = x_{n_x-1}$, $x_0 = x_{n_x}$, $x_{n_x+1} = x_1$, $F=5$. The simulation run of~\eqref{Lorenz96} is based on discretization using the explicit fourth-order Runge-Kutta method.  The discret-time system with process noise $w$ is expressed as:
\begin{align*} \label{Lorez96-discrete}
x_{j,k+1} = x_{j,k} + \frac{\Delta t}{6} \big(h_1 + 2h_2 + 2h_3 + h_4\big) + w_{j,k},
\end{align*}
where
\begin{align*}
h_1 &= f\left(x_{j,k}, F\right), \quad   h_2 = f\left(x_{j,k} + \Delta t \cdot h_1/2, F\right),\\
h_3 &= f\left(x_{j,k} + \Delta t\cdot h_2/2,  F\right),  \ h_4 = f\left(x_{j,k} + \Delta t\cdot h_3, F\right),
\end{align*}
$w_{j,k}\sim \mathcal{U}(-0.5, 0.5)$ is added as the i.i.d. process noise,  and $\Delta t = 0.01$ is the discretization time-step. In addition, we consider a nonlinear measurement function that partially observes 
every other component of the state:
\begin{align*}
y_{l,k} = x_{2l-1,k}  + \sin(x_{2l-1,k} ) + v_{l,k} ,
\end{align*}
where $l=1,\ldots,n_y = 20$, and $v_l \sim \mathcal{U}(-0.5, 0.5)$. We evaluate the state estimation performance of the E-IPF and U-IPF for the above system by the root mean squared error (RMSE) metric over 50 Monte Carlo (MC) runs. Each run is initialized randomly and with a bias to the true state. Meanwhile, we run the IPF based on numerical iteration (I-IPF) in~\cite{Chorin:ESAIM:2012}, EPF, and UPF for comparison. The PDF of the reference random variable $p(\xi_k)$ for all the considered IPFs is set to $\mathcal{N}(0, \alpha I)$, where $\alpha = 0.05$. Further, the individual Kalman filters use the covariances of $w$ and $v$ as $Q$ and $R$, respectively, in their estimation run.
\begin{figure}[t]\centering
\includegraphics[width=0.49\textwidth, trim={4cm 8.6cm 4.5cm 9cm},clip]{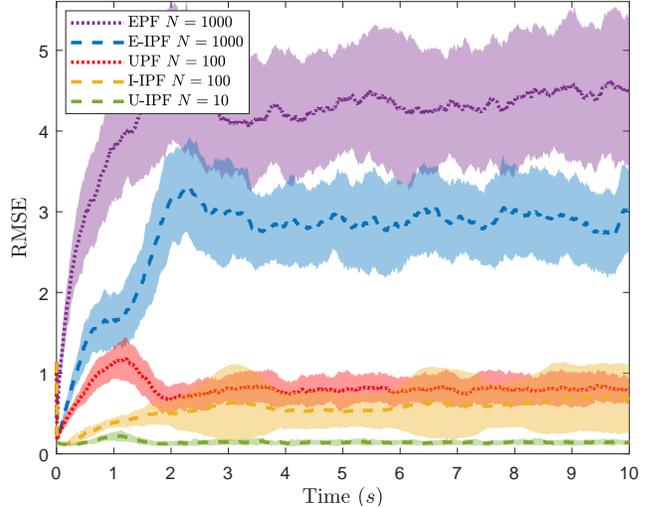}
\caption{RMSE performance of  the EPF/E-IPF (with 1,000 particles) and U-IPF (with only 10 particles) versus UPF/I-IPF (with 100 particles). The shaded regions correspond to the $\pm \sigma$ bounds by the filters over 50 Monte Carlo runs.}
\label{fig:RMSE}
\end{figure}

Fig.~\ref{fig:RMSE} shows the estimation performance of the different filters using different numbers of particles. The observations are summarized as follows. First, the KF-IPF algorithm provides advantageous estimation. Specifically, the E-IPF performs better than the EPF in terms of accuracy, with both using 1,000 particles; further, the U-IPF with only ten particles delivers considerably higher accuracy than the UPF with 100 particles. Second, compared to the I-IPF, the KF-IPF represents a more favorable and promising approach. Even though the I-IPF is understandably more accurate than the E-IPF (since it is based on iterative linearization and solution rather than the first-order linearization involved in the E-IPF), the U-IPF consistently outperforms the I-IPF, and this comes at using just ten particles rather than 100 by the I-IPF. The observations also suggest the need to use a good KF to well execute the KF-IPF algorithm. The simulation was run using MATLAB R2020b on a workstation equipped with a 3.5GHz Intel Core i9-10920X CPU, 128GB of RAM. Table~\ref{table:1} summarizes the average computation time of a MC run for the considered filters. Overall, the computation time increases with the number of particles. The U-IPF and I-IPF are computationally the fastest among the filters. The simulation results indicate the advantages and promise of the KF-IPF algorithm for nonlinear non-Gaussian estimation.
 \begin{table}[t]\caption{Average Computation Time}\label{table:1}
\setlength{\tabcolsep}{4pt}
\centering
        \begin{tabular}{@{}lllllll@{}}
        \toprule
        & EPF  & E-IPF &  UPF & I-IPF & U-IPF\\
	\midrule
	$N$ & 1,000 & 1,000 & 100 & 100 & 10 \\
	\midrule
        Time (s) & 3,462 &3,362 & 652 & 578.9 & 46.4 \\
        \bottomrule
        \end{tabular}
\end{table}
\vspace{-3pt}
\section{Conclusion}
The IPF has emerged as a vibrant tool for dealing with particle degeneracy confronting the PF. However, the present IPF methods are computationally cumbersome, which may limit their application. In this paper, we proposed approximately implementing the IPF framework {\em via} nonlinear KFs for higher computational tractability. We presented implicit importance sampling and then, on this basis, derived the IPF realization based on a bank of nonlinear KFs. We discussed the EKF and UKF realizations of the IPF. Further, we examined how the proposed realization relates to some existing PF methods. Finally, we demonstrated the effectiveness of KF-based IPF through the use of a simulation.

\section*{Acknowledgments}

The authors would like to thank the anonymous reviewers for their constructive comments. I. Askari and H. Fang were partially supported by the U.S. National Science Foundation under Awards CMMI-1763093 and CMMI-1847651.   X. Tu was partially supported by the U.S. National Science Foundation under Award DMS-1723066.

 \balance
\bibliographystyle{elsarticle-num}

\bibliography{cit}

\end{document}